\documentclass[sigconf]{acmart}

\AtBeginDocument{%
 }

\copyrightyear{2023}
\acmYear{2023}
\setcopyright{rightsretained}
\acmConference[CIKM '23]{Proceedings of the 32nd ACM International Conference on Information and Knowledge Management}{October 21--25, 2023}{Birmingham, United Kingdom}
\acmBooktitle{Proceedings of the 32nd ACM International Conference on Information and Knowledge Management (CIKM '23), October 21--25, 2023, Birmingham, United Kingdom}\acmDOI{10.1145/3583780.3615500}
\acmISBN{979-8-4007-0124-5/23/10}

\makeatletter
\gdef\@copyrightpermission{
  \begin{minipage}{0.3\columnwidth}
   \href{https://creativecommons.org/licenses/by/4.0/}{\includegraphics[width=0.90\textwidth]{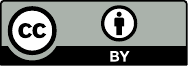}}
  \end{minipage}\hfill
  \begin{minipage}{0.7\columnwidth}
   \href{https://creativecommons.org/licenses/by/4.0/}{This work is licensed under a Creative Commons Attribution International 4.0 License.}
  \end{minipage}
  \vspace{5pt}
}
\makeatother

\usepackage{booktabs}
\usepackage{multirow}
\usepackage{stfloats}
\usepackage{graphicx}
\usepackage{subfigure}
\usepackage[export]{adjustbox}
\usepackage{bbm}
\usepackage{amsmath}

\begin{document}

\title{SPM: Structured Pretraining and Matching Architectures for Relevance Modeling in Meituan Search}

\author{Wen Zan}
\email{zanwencs@gmail.com}
\affiliation{%
  \institution{Meituan Inc.}
  \streetaddress{4 East Wangjing Road}
  \city{Beijing}
  \country{China}
}

\author{Yaopeng Han}
\email{hanyaopeng95@163.com}
\affiliation{%
  \institution{Meituan Inc.}
  \streetaddress{4 East Wangjing Road}
  \city{Beijing}
  \country{China}
}

\author{Xiaotian Jiang}
\authornote{Corresponding Author.}
\email{jiangxiaotian.work@gmail.com}
\affiliation{%
  \institution{Meituan Inc.}
  \streetaddress{4 East Wangjing Road}
  \city{Beijing}
  \country{China}
}

\author{Yao Xiao}
\email{xiaoyao06@meituan.com}
\affiliation{%
  \institution{Meituan Inc.}
  \streetaddress{4 East Wangjing Road}
  \city{Beijing}
  \country{China}
}

\author{Yang Yang}
\email{yangyang113@meituan.com}
\affiliation{%
  \institution{Meituan Inc.}
  \streetaddress{4 East Wangjing Road}
  \city{Beijing}
  \country{China}
}

\author{Dayao Chen}
\email{chendayao@meituan.com}
\affiliation{%
  \institution{Meituan Inc.}
  \streetaddress{4 East Wangjing Road}
  \city{Beijing}
  \country{China}
}

\author{Sheng Chen}
\email{chensheng19@meituan.com}
\affiliation{%
  \institution{Meituan Inc.}
  \streetaddress{4 East Wangjing Road}
  \city{Beijing}
  \country{China}
}

\renewcommand{\shortauthors}{Wen Zan et al.}

\begin{abstract}

In e-commerce search, relevance between query and documents is an essential requirement for satisfying user experience. Different from traditional e-commerce platforms that offer products, users search on life service platforms such as Meituan mainly for product providers, which usually have abundant structured information, e.g. name, address, category, thousands of products. Modeling search relevance with these rich structured contents is challenging due to the following issues: (1) there is language distribution discrepancy among different fields of structured document, making it difficult to directly adopt off-the-shelf pretrained language model based methods like BERT. (2) different fields usually have different importance and their length vary greatly, making it difficult to extract document information helpful for relevance matching. 

To tackle these issues, in this paper we propose a novel two-stage pretraining and matching architecture for relevance matching with rich structured documents. At pretraining stage, we propose an effective pretraining method that employs both query and multiple fields of document as inputs, including an effective information compression method for lengthy fields. At relevance matching stage, a novel matching method is proposed by leveraging domain knowledge in search query to generate more effective document representations for relevance scoring. Extensive offline experiments and online A/B tests on millions of users verify that the proposed architectures effectively improve the performance of relevance modeling. The model has already been deployed online, serving the search traffic of Meituan for over a year.

\end{abstract}



\maketitle

\section{Introduction}

Unlike traditional e-commerce services that mainly provide products \cite{jiang2019unified,nigam2019semantic}, most users on life service e-commerce platforms , like Meituan \footnote{https://meituan.com/}, would like to search for service providers instead, such as restaurants, bars, stores, hotels, etc. These two forms of documents differ in that a service provider usually contains a great number of structured information therein. For example, a restaurant may have structured fields like name, address, category, comments, tags, and services like dishes and coupons. Considering all theses different types of structured contents in relevance matching is a challenging task. 

To model the semantic relevance between query and document, previous methods tend to use a two-stage training paradigm \cite{jiang2020bert2dnn}. At the first stage, a relevance model is first pretraining on a large-scaled dataset with self-supervised tasks. Afterwards, the model is further trained on a downstream dataset for task-specific finetuning. 

However, previous e-commerce relevance models are less effective  for documents with rich structured contents \cite{yao2021learning}. 

Specifically, at pretraining stage, existing relevance methods are normally designed for product search, and each document has only limited fields \cite{zhang2020bert}. However, in life service scenario, there are various fields in each document, and each field may have its own structure. Under this circumstance, it is impractical to directly input all the contents in a document for pretraining model, let alone the distributional discrepancy among different fields. 

At finetuning stage, the models used in relevance task can be categorized into two architectures: cross-encoder and bi-encoder. Cross-encoder models better capture the interactions between query and document, and thus offer better performance. 
In e-commerce search scenario, bi-encoder models are widely used, since this architecture can cache the representations of query and document, and thus reduce calculations at online inference stage \cite{liu2021que2search}. However, bi-encoder methods usually get lower performance compared to cross-encoder, due to the lack of interactions between query and document.

In this paper, we propose novel pretraining and matching architectures for relevance modeling in e-commerce search, especially for documents with rich structured contents. First, based on the characteristics of relevance task and document structure, we propose an effective pretraining method that employs both query text and multiple fields of document as inputs. Specifically, this method first extract basic fields of a document, such as name, brand, category, address, etc. Then, for lengthy fields like coupons or dishes, a field compression method is designed to effectively extract the pivotal information therein. Considering the vocabulary gap between query text and document, we also add document-related query to make the pretrained model better understand query. 

After that, we design different mask strategies for each field to better adopt the word distribution of their own. Experiment shows that the proposed pretraining method brings better performance on relevance scoring with cross-encoder models, proving the effectiveness of modeling the rich structured information in documents. However, when we use bi-encoder model instead, the performance was not improved, and even worse than its original version that use query and name field only. It means that the document representation obtained by this model cannot effectively capture the rich information contained in the document. 
To solve this problem, in this paper we use query intent signals in search engine and design information extractors to generate better document representation. This method enables the query and document information to interact earlier, i.e, in the process of generating document representation. 
Therefore, it can effectively extract the information most beneficial for relevance matching.

To summarize, the contributions of this paper are as follows:
\begin{itemize}
\item We propose an effective multi-field pretraining architecture for e-commerce search with rich structured documents, and design a field compression algorithm to extract pivotal information from lengthy fields.
\item We propose a novel matching architecture for structured documents with effective document extractors to generate more informative document representations for relevance scoring, utilizing domain knowledge in search engines.
\item We conduct extensive experiments to evaluate the effectiveness of the proposed methods. Results on offline evaluation and online A/B test show that the methods can significantly improve the performance of relevance scoring in e-commerce search.
\end{itemize}

\section{Related Works}
\subsection{Pretraining Models}
In recent years, pretrained models like BERT \cite{devlin2018bert}, RoBERTa \cite{liu2019roberta}, XLNet \cite{yang2019xlnet} have brought great performance improvements on many NLP tasks. These models are first pretrained on large-scale unsupervised data, and then fine-tuned on task-specific data for downstream tasks. BERT is a representative work among them. It obtains text representations by using transformer \cite{vaswani2017attention} architecture and training on two tasks - MLM (Mask Language Model) and NSP (Next Sentence Prediction). Recently, \cite{yang2019xlnet,liu2019roberta} prove that the NSP task is not indispensable.

Since pretrained models usually do not model knowledge of specific downstream tasks, many works have begun to study how to encode domain knowledge into these models. Gururangan \textit{et al.} \cite{gururangan2020don} pointed out that continuing pretraining with domain data can gain performance improvements in domain-specific downstream tasks. Zou \textit{et al.} \cite{zou2021pre} proposed a pretraining paradigm for relevance modeling in web search. Zhang \textit{et al.} \cite{zhang2021ace} proposed a pretraining work incorporating multi-modal information in e-commerce scenarios. Zhang \textit{et al.}  \cite{zhang2020bert} proposed a e-commerce pretrained model with phrase mask and document neighbor for augmentation. Despite the success these models made in their domains, there is still lack of pretraining methods designed for documents with rich structured contents, which is very common in life service e-commerce scenario.

\subsection{Text Matching}
Text matching (or text relevance) task computes the similarity between two sentences. In the recent decade, neural methods have been flourishing for their better capability to model semantic similarity. Neural text matching methods can be categorized into two types: cross-encoder methods \cite{pang2016text,mitra2017learning,dai2018convolutional,wu2016sequential}, and bi-encoder methods \cite{hu2014convolutional,huang2013learning,neculoiu2016learning,wan2016deep}. They differs in that the former type concatenates two sentences together before feeding them into the model to calculate similarity, while the latter one first gets representations for each of the two sentence before calculating similarity score between them. The advantage of cross-encoder methods lies in that information of the two sentences can be well interacted, so these models normally have better performance. The advantage of bi-encoder models lies in that the trained sentence representations can be offline cached, which greatly speeds up online inference of the model. 
Recently, BERT has shown its superiority on text matching task. To increase interactions between sentences for bi-encoder BERT methods, Reimers \textit{et al.} \cite{reimers2019sentence} proposed Siamese BERT networks that use a shared transformer for both of the sentences.  Khattab \textit{et al.} \cite{khattab2020colbert} proposed ColBERT that uses a sum-of-max operator to facilitate interaction between representations of the two sentences. Humeau \textit{et al.} \cite{humeau2019poly} use additional parameters to extract multiple representations of the longer sentence. However, these methods treat two sentences as plain text, while in industrial search engines both side of the match have rich additional knowledge to be used.

\section{Methodology}
In this section, we demonstrate the details of the proposed methods.
First, we introduce the pretraining model designed for structured document (\textbf{S}tructured \textbf{P}retraining BERT, i.e.,SPBERT), including two parts: data construction and pretraining strategy.
Then we introduce the matching architecture for structured information, and propose two information extractors to better match query and structured documents in e-commerce scenario.

\begin{figure}[tp]
    \centering
    \includegraphics[width=0.9\linewidth]{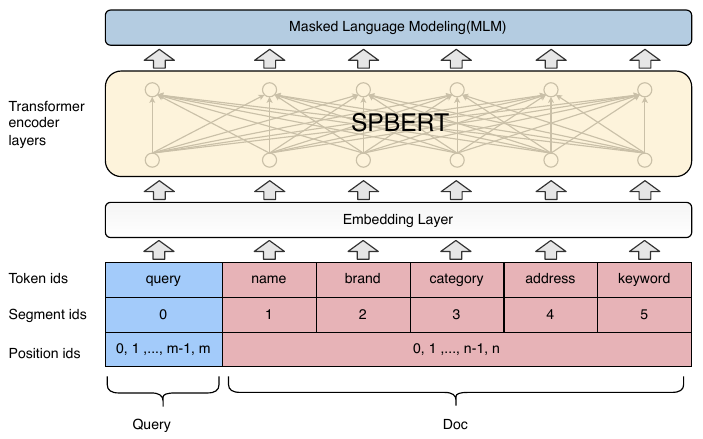}
    \caption{Structured Pretraining BERT (SPBERT) model.}
    \label{fig:pretrain}
\end{figure}

\subsection{Pretraining Model for Structured Document}

We formally define the $i_{th}$ document with structured information as 
$D_i=\{\mathbb{W}_{i}^{F_1},\mathbb{W}_{i}^{F_2},\cdots,\mathbb{W}_{i}^{F_K}\}$, where $K$ denotes the number of input fields.
Let $\mathbb{W}_{i}^{F_k}=\{w_{i1}^{F_k}, w_{i2}^{F_k}, \cdots, w_{i{\vert F_k\vert}}^{F_k}\}$ denotes all tokens in the $k_{th}$ field in $D_i$, $w_{ij}^{F_k}$ denotes the $j_{th}$  token in $\mathbb{W}_{i}^{F_k}$, and $|F_k|$ denotes the number of tokens in $\mathbb{W}_{i}^{F_k}$.

Considering the fact that fields (such as name, brand, address, category, products, etc.) of a document in e-commerce scenario are normally not independent of each other, we define the following objective function for pretraining:
\begin{small} 
\begin{align}
\mathop{max}\limits_{\theta } \,  \sum_{i=1}^{\vert D\vert}\sum_{k=1}^{K}\sum_{j=1}^{\vert F_k\vert}   \, m_{ij}^{F_k}  \log {p_\theta }(& w_{ij}^{F_k}| \mathbb{W}_{i}^{F_1},\cdots ,\mathbb{W}_{i}^{F_{k-1}} ,\cdots, \mathbb{W}_{i}^{F_{k+1}}, \cdots,\mathbb{W}_{i}^{F_{K}}, \nonumber\\
 & w_{i1}^{F_k},\cdots,w_{i(j-1)}^{F_k}, w_{i(j+1)}^{F_k},\cdots,w_{i\vert F_k \vert}^{F_k})
\end{align}
\end{small}
where $m_{ij}^{F_k} \geq 0$  denotes how many times token $w_{ij}^{F_k}$ is sampled to be masked for MLM\cite{devlin2018bert} pretraining task.

The model structure is illustrated in Figure \ref{fig:pretrain}. In order to distinguish different fields which are concatenated together as input, we use a separate segment embedding for each of them. Besides, we use position embedding starting from 0 for query and document respectively, so that the words in the same position of query and name will get the same position embedding.

In this section, we introduce the proposed matching framework with structured information extractors.

\subsubsection{Matching Framework}
\label{chapter:mfwk}
\begin{figure}[tp]
    \centering
    \includegraphics[width=0.9\linewidth]{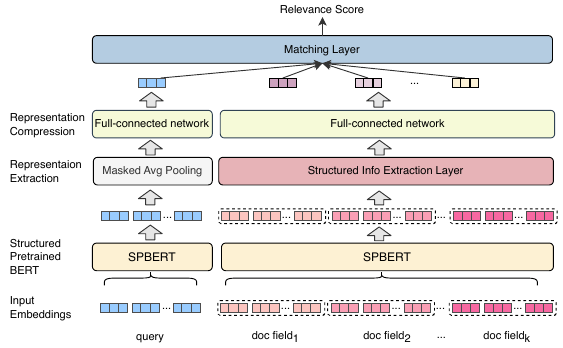}
    \caption{Matching framework}
    \label{fig:matching}
\end{figure}

As depicted in Figure \ref{fig:matching}, the framework uses a bi-encoder architecture to encode tokens of query and document respectively and obtains their representations. The advantage of bi-encoder architecture is that the output embeddings of the document can be calculated and cached in advance to speed up online serving \cite{yao2021learning}. 
A representation extraction layer is designed to extract the most important information beneficial for matching from the sophisticated structured input. Details of the structured information extractors are elaborated in section \ref{chapter:ige}.
After that, a representation compression layer is adopted to reduce dimension of the extracted representations. 
Then, relevance score between query and doc can be calculated through the matching layer. 

\begin{figure*}[htpb]
    \centering
    \includegraphics[width=0.8\linewidth]{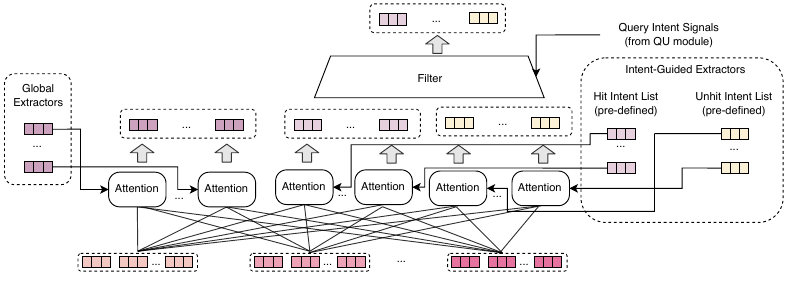}
    \caption{Illustration of the proposed intent-guided information extractor.}
    \label{fig:ige}
\end{figure*}

\subsubsection{Intention-Guided Extractor(IGE)}
\label{chapter:ige}
In this section, we further propose an intention-guided information extractor, as depicted in Figure \ref{fig:ige}. It utilizes the intention signal of the query to be matched in advance, and extracts information according to the guidance of this knowledge. Specifically, we use the \textbf{q}uery \textbf{u}nderstanding (QU) module in search engine to provides query intention signal, and then extract the information in the structured document according to the signal.

Formally, a group of intention-guided extractors $h^{I_1}, \cdots, h^{I_M}$, $ h^{I_1^\prime }, \cdots, h^{I_M^\prime}$ is used to extract query intention specific representation from document with $K$ fields. Here, $M$ denotes the total number of intention types, $h^{I_i}$ and $h^{I_i^\prime}$ respectively denotes the corresponding extractors that the $i_{th}$ intention is hit or not, $h_{i}^{F_k}$ denotes the representation after SPBERT encoder of the $i_{th}$ token from $k_{th}$ field. The resultant representations of extractor $h^{I_i}$ and $h^{I_i^\prime}$ are denoted as $\hat{h}^{I_i}$ and $\hat{h}^{I_i^\prime}$ respectively, which are defined as follows:
\begin{equation}\label{att-ige}
  \begin{split}
[\hat{h}^{I_i}, \hat{h}^{I_i^\prime}  ]=  \textit{attention}([h^{I_i}, h^{I_i^\prime}], [h_{1}^{F_1},\cdots,h_{\vert F_K \vert}^{F_K} ])
  \end{split}
\end{equation}
Where $\textit{attention}(X, Y) = \textit{softmax}(XY^T)Y$. Additionally, we also use a group of $T$ global extractors $[h^{G_1}, \cdots, h^{G_T}]$ to extract information extract global document representations $[\hat{h}^{G_1} , \cdots, \hat{h}^{G_T}]$ in the same way in Eq.\ref{att-ige}. The output of the intention-guided extractor layer is $[\hat{h}^{I_1},\cdots, \hat{h}^{I_M} ,\hat{h}^{I_1^\prime},\cdots,\hat{h}^{I_M^\prime}, \hat{h}^{G_1} , \cdots, \hat{h}^{G_T} ]$.

Then, a compression layer is utilized to obtain a memory-friendly version of representation for deployment. Specifically, we adopt a full-connected network to project the extracted representation from a high dimension to a much lower dimension. For simplicity, we use the same notations for representations before and after compression layer. Note that these are embeddings we will pre-calculate and cache when we deployment the model. 

In matching layer, we first index out intention related representations from $[\hat{h}^{I_1},\cdots, \hat{h}^{I_M} ,\hat{h}^{I_1^\prime},\cdots,\hat{h}^{I_M^\prime}]$ according to the corresponding intention signals, indicate each predefined intention is hit or not hit by the input query, obtained from query understanding module in search engine. The selected intention related representations are noted as $[\hat{h}^{\tilde{I}_1},\cdots, \hat{h}^{\tilde{I}_M}]$. Note that although we will cache $2*M+T$ embeddings for each document, only $M+T$ embeddings are involved during matching. By aggregating these representations using attention, we get the final document representation:
\begin{equation}
  \begin{split}
  [\bar{h}^d ]= \textit{attention}([h^q], 
 [\hat{h}^{\tilde{I}_1},\cdots, \hat{h}^{\tilde{I}_M}, \hat{h}^{G_1} , \cdots, \hat{h}^{G_T} ])
  \end{split}
\end{equation}
Where $h^q$ is the query representation after the compression layer.

At last, cosine similarity is used to calculate the relevance score between query representation and document representation:
\begin{equation}
    s={\textit{cosine}(h^q, \bar{h}^d)}
\end{equation}

We use the following loss function to characterize relevance at a finer granularity, since the label is multi-leveled:
\begin{equation}
\label{eq:loss}
     \textit{loss} =  max(0, (s-y)^2-\epsilon)
\end{equation}
where $y\in [0,1]$ denotes the normalized relevance label. $\epsilon$ is the margin hyper-parameter, which is set to 0.1 in our experiments.

\section{Experiment}
To demonstrate the effectiveness of the proposed method, we conduct extensive experiments on an industrial life service e-commerce search engine. 
Experimental results show that for e-commerce search with rich structured documents, the proposed pretraining and matching architectures can significantly improve the performance of relevance modeling.

\subsection{Datasets}

\subsubsection{Pretraining Dataset}
The dataset at this stage is mainly composed of structured documents from search engine results. Specifically, we sampled 50 million documents from Meituan search, a large-scale real-world life service e-commerce search engine, to build the training corpus. These documents include fields like name, brand, category, address, keywords. In addition, for each frequency-visited document, we sample a top query and add it as a new field to the training corpus to better adapt to the relevance task.

\subsubsection{Relevance Dataset}
\label{chapter:dataset}
The relevance dataset consists of more than 1 million human-labeled query-document pairs from historical search results from Meituan Search. With the help of Meituan’s crowd-sourcing platform, each of these query-document pairs is annotated as one of the three relevance levels: strong relevant, weak relevant, not relevant.  Specifically, there are
1,068,135 strong relevant samples, 169,454 weak relevant samples, and 249,695 irrelevant samples. 
The labels $y$ used for the strong relevant, weak relevant, and irrelevant samples in Equation (\ref{eq:loss}) are $1$, $0.7$, and $0$, respectively.
We sample 10,000 samples for devset and 10,000 for testset to evaluate models, and the rest 1,467,284 samples are used for training model.

\subsection{Evaluation Metrics}
We use the following evaluation metrics to evaluate the performance of the proposed method.

The Area Under Curve(\textbf{AUC}) is widely used as
the evaluation metric in e-commerce scenarios\cite{jiang2019unified,yao2021learning} for evaluating relevance model . If the relevance score given by the model is ideally all higher on the relevant samples than the irrelevant ones, AUC will get the maximum value of 1. Traditional AUC formula is only suitable for binary classification, but our relevance matching task has three relevance levels. So we employ a multiclass AUC formula:
\begin{equation}
    \textit{AUC} = \frac{1}{N} \sum_{i}^N \frac{\sum_j\sum_k \mathbbm{1}(y_{ij}>y_{ik})\cdot \mathbbm{1}(s_{ij}>s_{ik})}{\sum_j\sum_k \mathbbm{1}(y_{ij}>y_{ik})}
\end{equation}
where $y_{ij}\geq 0$ is the relevant label for the $j_{th}$ returned document of the $i_{th}$ query, $s_{ij} \in [0,1]$ is the model score for the $j_{th}$ returned document of the $i_{th}$ query and $N$ is the total number of test queries.

The \textbf{Badcase@5} is an important metric we used to evaluate the quality of the top search results of the online system. Specifically, it calculates the proportion of queries having irrelevant cases in the top 5 ranking results:
\begin{equation}
    Badcase@5 = \frac{1}{N} \sum_{i}^N \mathbbm{1}(\sum_j^5 y_{ij}=0)
\end{equation}

\subsection{Competitor System}
To evaluate the effect of the proposed structured pretraining method on relevance task, we compare the following \textbf{C}ross-\textbf{E}ncoder methods. We use $^+$ to denote that the models use a query and multiple fields of a document as inputs. Otherwise, the inputs consist only of the query and the name field of the document.
\begin{itemize}
    \item \textbf{BERT-Large-CE}: a cross-encoder matching model based on BERT-Large (24 layers) with query and document name as input.
    \item \textbf{BERT-Large-CE$^+$}: the same model structure with BERT-Large-CE, but with query and structured document as input. Fields of the document are concatenated with [SEP], using the same segment embedding.
    \item \textbf{BERT-CE$^+$}: a cross-encoder matching model based on a 6-layer BERT model distilled from BERT-Large, using query and structured document as input.
    \item \textbf{SPBERT-Large-CE$^+$}: a cross-encoder matching model based on SPBERT-Large, the 24-layer structured pretraining model. Using query and structured document as input.
    \item \textbf{SPBERT-CE$^+$}: a cross-encoder matching model based on a 6-layer SPBERT model distilled from SPBERT-Large, using query and structured document as input.
\end{itemize}

To improve model efficiency on online serving, relevance models in e-commerce scenario often adopt \textbf{B}i-\textbf{E}ncoder architecture. This architecture can cache representations of queries and documents in advance and thus reduce online calculation. We further evaluate the performance of the pretraining model by comparing it with methods of this architecture.
\begin{itemize}
    \item \textbf{BERT-BE}: a bi-encoder matching model based on BERT with query and document name as input.
    \item \textbf{SPBERT-BE}: a bi-encoder matching model based on SPBERT with query and document name as input.
    \item \textbf{SPBERT-BE$^+$}: the same model structure with SPBERT-BE with query and structured document as input.
\end{itemize}

To evaluate the effectiveness of the proposed matching methods with structured extractors, we compare them with state-of-the-art late interaction methods for comparison.
\begin{itemize}
    \item \textbf{SPBERT-SBERT$^+$}: SBERT \cite{reimers2019sentence} architecture using SPBERT as pretrained model. As one of late interaction methods, SBERT concatenate input query embedding, document embedding, and simple mathematical operation result between them before maching. In this way, this method usually gets better results compared with other methods directly matching query embedding and document embedding.
    \item \textbf{SPBERT-ColBERT$^+$}: ColBERT \cite{khattab2020colbert} architecture using SPBERT as pretrained model. Since ColBERT model calculates the matching score of each token of a query to all tokens of a document, it tends to gain better performance than other bi-encoder methods. However, this method needs to cache the token representations of all documents beforehand, which is memory expensive and hard to deploy when documents have long length.
    \item \textbf{SPBERT-PolyEncoder$^+$}: PolyEncoder \cite{humeau2019poly} architecture using SPBERT as pretrained model. PolyEncoder uses multiple codes to interact with document to obtain effective document representations. By controlling the number of codes, this method can flexibly trade off between memory usage and model performance.

    \item \textbf{SPBERT-IGE$^+$}: a matching method with the intent-guided extractor elaborated in section \ref{chapter:ige}, using SPBERT as pretrained model. This method extracts the information of structured documents for all types of intent signals.
\end{itemize}

\subsection{Experimental Setting}
For all the experiments, we set the learning rate to 2e-5, warm-up ratio to 10\%, dropout rate to 0.1, and use the Adam \cite{kingma2014adam} optimizer. 

In the pretraining stage, the model parameters are set to 24 hidden layers, 16 attention heads, hidden size of 1024, and feed-forward layers with dimension 4096; the batch size is set to 20 during training, and the model is distributedly trained on 64 NVIDIA V100 GPUs for 10 epochs.

In the distillation stage, we use the pretraining model as the teacher. The parameters of the student model are set to 6 hidden layers, 12 attention heads, a hidden size of 384, and feed-forward layers with dimension 1200. During training, the batch size is set to 8 and the model is distributedly trained on 32 NVIDIA V100 GPUs for 3 epochs.

In the finetuning stage for relevance task, we set the sequence length to 160 for cross-encoder models. As to the bi-encoder models, we set the maximum query length to 32, and the maximum document length to 128. Before the matching layer, we use a one-layer fully-connected network as a compression layer to project the dimension of embeddings from 384 to 32. For SPBERT-IGE$^+$, the number of intent-guided extractors is set to 3 corresponding to the number of query intent types we use, and we adjust the number of global extractors to meet the requirements on the total number of extractors in each experiment. The embeddings of all extractors are random initialized before training. $\epsilon$ in equation (\ref{eq:loss}) is set to $0.01$. Batch size is set to 256 during training. Early stop is performed when AUC does not improve within 5 epochs.

To the best of our knowledge, there is no public e-commerce relevance dataset that has highly structured documents as well as query intent signals. 
Therefore, we report the evaluation results of the proposed method on the dataset described in \ref{chapter:dataset} for all offline experiment.

\subsection{Offline Experimental Results}
Table \ref{tab:pretrain} reports the experimental results of the base model BERT-Large, the structured pretraining model SPBERT-Large, as well as the corresponding distilled models BERT and SPBERT, using all fields or name field as document input respectively. Comparing BERT-Large-CE with BERT-Large-CE$^+$, we can observe that  adding structured inputs to unstructured pretraining models is harmful for their performance. This may be attributed to the distributional discrepancy among different fields of structured document, which can not be distinguished by unstructured pretraining models. On the contrary, the comparison between BERT-Large-CE and SPBERT-Large-CE$^+$ informs us that using structured input with structured pretraining model does not lower model performance, but instead improves AUC by as much as 1.8\%. This result reveals that adding structured information is beneficial for relevance tasks. Moreover, this benefit can only be achieved by using structured pretraining. Finally, by comparing the distilled model BERT-CE and SPBERT-CE$^+$, it can be seen that the advantages of structured pretraining are still preserved after distillation.

To improve model efficiency, e-commerce relevance methods often use bi-encoder architecture to reduce real-time computation. Table \ref{tab:pretrain} reports the performance of BERT and SPBERT of this form. First, comparing SPBERT-BE with BERT-BE, it can be observed that SPBERT can still achieve better results even using unstructured input. This is because the term-weight-based masking strategy employed on name field at pretraining stage enables the words important to the matching task to be better learned. Note that by comparing SPBERT-BE with SPBERT-BE$^+$, we can see that the performance of bi-encoder model decreases when using structured input. This is due to the fact that structured document includes multiple fields with different importance to the relevance task, so it is difficult to extract enough effective information by just using pooling. This observation inspires us to design the more effective information extraction methods based on the characteristics of structured document.

\begin{table}[H]
\renewcommand\tabcolsep{15pt} 
\caption{Performance of structured pretraining models on relevance task }
\label{tab:pretrain}
\centering
\begin{tabular}{l|c} 
\toprule
\textbf{Model}                & \textbf{AUC(\%) } \\ 
\midrule
BERT-Large-CE        & 87.88      \\
BERT-Large-CE$^+$   & 86.48      \\
SPBERT-Large-CE$^+$ & 89.60       \\ 
\hline
BERT-CE              & 86.52      \\
SPBERT-CE$^+$       & 88.51      \\
\hline
BERT-BE              & 82.19      \\
SPBERT-BE       & 82.64      \\
SPBERT-BE$^+$       & 81.91      \\
\bottomrule
\end{tabular}
\end{table}

To prove the effectiveness of the proposed matching architectures, we compare them with state-of-the-art late interaction bi-encoders. In consideration of fairness, the number of codes in PolyEncoder and extractors in SPBERT-IGE$^+$ are all set to 8. As shown in Table \ref{tab:finetune}, late interaction methods alleviate the problem of performance degradation after adding structured information in bi-encoder. Moreover, we observe that SPBERT-IGE$^+$ has better performance than SPBERT-SBERT$^+$ and SPBERT-PolyEncoder$^+$. This indicates the proposed  extractor can better extract informative representations for matching than those that extract information completely on the model itself. Note that SPBERT-
IGE$^+$ has reached close performance to SPBERT-ColBERT$^+$, while it at the same time needs much fewer document representations that need to be cached than the latter, meaning that the proposed method is both effective and memory-friendly.

\begin{table}[H]
\renewcommand\tabcolsep{1pt} 
\caption{Comparison of the proposed matching methods.}
\label{tab:finetune}
\begin{tabular}{l|c|c}
\toprule
\multirow{1}{*}{\textbf{Model}}       & \multirow{1}{*}{\textbf{AUC(\%)}} & \textbf{Doc Embedding Storage} \\
\midrule
SPBERT-BE                             & 82.64	& 1x                      \\
SPBERT-BE$^+$          & 81.91	& 1x                      \\
SPBERT-SBERT$^+$     & 82.37	& 1x                   \\
SPBERT-ColBERT$^+$     & 85.46	& 128x                   \\
SPBERT-PolyEncoder$^+$ & 84.93	& 8x                      \\
\textbf{SPBERT-IGE$^+$}         & 85.37	& 11x (8x is used during matching)       \\             
\bottomrule
\end{tabular}
\end{table}

\subsection{Online Experimental Results}

\subsubsection{Deployment}

\begin{figure}[htbp]
    \centering
    \includegraphics[width=1.\linewidth]{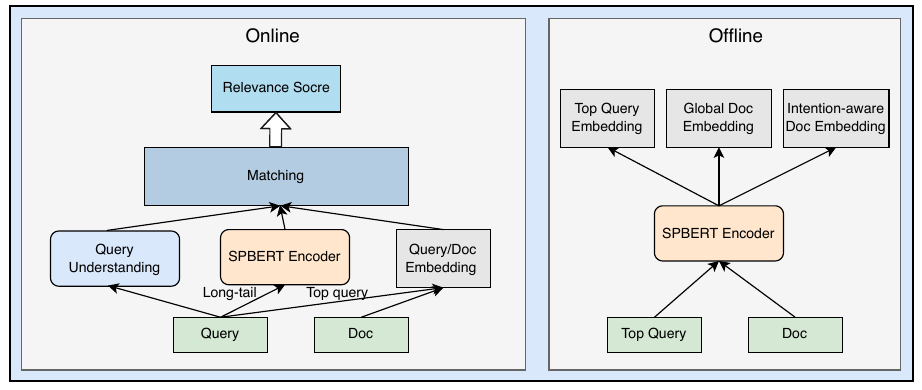}
    \caption{Deployment of the proposed relevance model}
    \label{fig:deployment}
\end{figure}

Meituan search faces tens of millions of queries every day, so the online system has strict restriction on latency. Since BERT uses multiple layers of transformers, its efficiency decrease as the number of layers increases. For the sake of efficiency, we deploy the distilled BERT for online serving. As illustrated in Figure \ref{fig:deployment}, to further lower the latency, we cache the extracted multiple embeddings of the documents offline. We also cache embeddings of the top queries offline, while calculate that of the long-tail queries online. The intent signal of queries are obtained from query understanding module in search engine. Then, the matching layer uses these results to calculate relevance score. Finally, this score is discretized into relevance level, and used by the search system for stratification, namely sorting the search results stably by the relevance level in descending order. 

\subsubsection{Online A/B Test}

We conduct an online A/B test for one week to ensure that our new proposed model (i.e.,SPBERT-IGE$^+$ ) will improve the system performance compared with the old one (i.e., BERT-BE). The results show that the new model can largely improve the overall user experience of the e-commerce search system. In particular, the Badcase@5 metric has decreased by $1.12\%$ which is statistically significant with $p < 0.05 $. This show that the structured pretrain and matching architectures with intent-guided extractors are very helpful for improving the relevance of the search system.

\section{Conclusion}
In this paper, we describe novel pretraining and matching architectures for relevance matching with rich structured documents. To make the method aware of the distributional discrepancy among fields and the different importance of each field to relevance task, we propose a structured pretraining architecture that uses multiple fields to pretrain a language model. To generate informative representations for relevance scoring, we propose a matching architecture with query intent guided document information extractors. Extensive offline and online experiments are conducted to verify the effectiveness of the proposed architectures and the corresponding model are successfully implemented and deployed on Meituan Search to boost search relevance.

\bibliographystyle{ACM-Reference-Format}
\balance
\bibliography{sample-base}

\clearpage

\appendix

\section{Ablation Study}

\begin{figure}[htbp]
  \centering
  \includegraphics[width=1\linewidth]{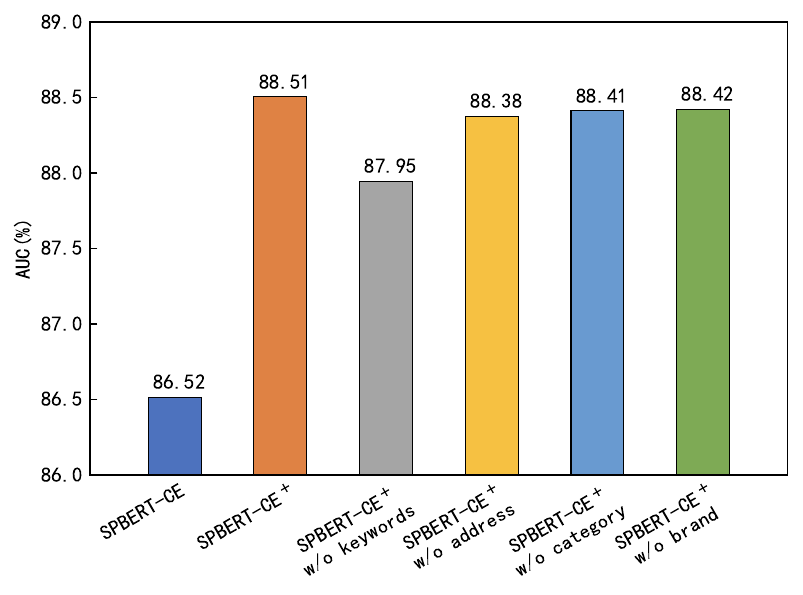}
  \caption{Importance of different document fields.}
  \Description{DESCRIPTIONS PLACEHOLDER.}
  \label{fig:fields}
\end{figure}

\subsection{Importance of Each Field in Structured Input}
To investigate the impact of each field in the structured pretraining model for relevance task, we compare the performance of the following variants: models with unstructured or structured input, and models that drop each field, e.g. category, keywords, address, brand. As shown in Figure \ref{fig:fields}, we can find out that each field has its contribution to the performance of the model, and the contribution of each field varies.

\begin{figure}[htbp]
  \centering
  \includegraphics[width=1\linewidth]{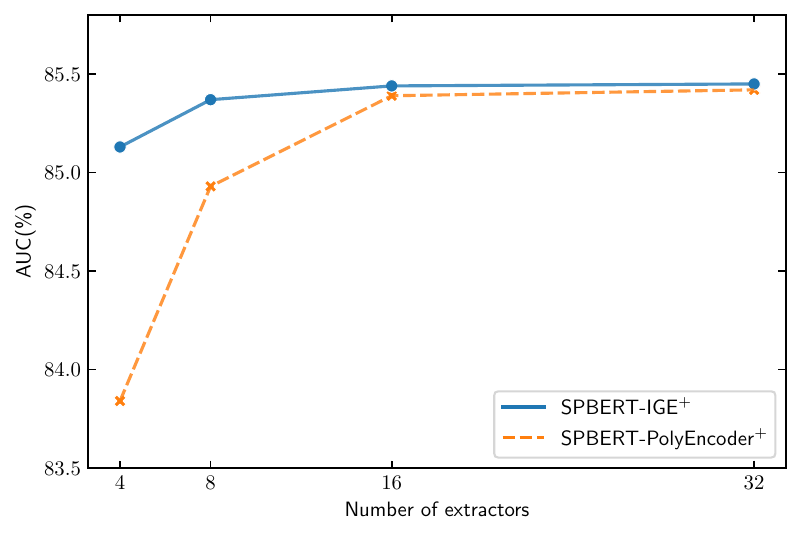}
  \caption{Impacts of different number of extractors.}
  \Description{DESCRIPTIONS PLACEHOLDER.}
  \label{desp}
\end{figure}

\subsection{Effect of the Number of Extractors}
We conduct experiments on how the number of extractors influences the performance of the proposed models. 
For PolyEncoder, we tune the number of extractors to reach the target number. For the proposed models, we tune the number of global extractors to make the number of all extractors reach the target number. The results shown in Figure \ref{desp} show that the performance of both models improves and finally achieves similar results as the number of extractors increases. 
However, the number of extractors cannot be increased infinitely, since the required  storage size to cache document embedding grows linearly as the number of extractors increase and there are usually tens of millions of documents in real search system. 
Note that SPBERT-IGE$^+$ significantly outperforms SPBERT-PolyEncoder$^+$ when the extractors number is relatively small(e.g. 4 or 8). In online system., we deploy the SPBERT-IGE$^+$ model with 8 extractors.

\end{document}